\documentclass{article}  
\usepackage{graphicx}
\usepackage{txfonts}
\usepackage{psfig}
\begin{document}
\noindent
{\em 2021 outburst of RS Oph  \hfill{2 March 2022}}
\begin{center}
\vskip 0.5 cm
{\large \bf The 2021 outburst of RS Oph: a pictorial atlas of the spectroscopic
evolution. II.} \\
\smallskip
{\large \bf From day 19 to 102 (solar conjunction)}\\
\vskip 0.7 cm
{\large Ulisse Munari$^1$ and Paolo Valisa$^2$}
\end{center}
\vskip 0.3 cm
\noindent
\phantom{~~~~~~~~~~~}{\small 1: INAF National Institute of Astrophysics, 36012 Asiago, Italy}\\
\phantom{~~~~~~~~~~~}{\small 2: ANS Collaboration, c/o Astronomical Observatory, 36012 Asiago, Italy}\\
\vskip 0.5 cm

\baselineskip 9pt
\begin{center}
\parbox{13.4cm}{\baselineskip 9pt
{\bf Abstract}. {\small
We present the second part of our atlas of the spectral evolution of RS Oph
during the 2021 nova outburst, characterized by a tight-monitoring in
high-resolution with Echelle spectrographs mounted on the Varese 0.84m and
Asiago 1.82m telescopes.  In this Paper~II we cover from day 19 to 102 (Aug 28 to Nov
19), the interval going from appearance of high ionization lines to the
stop imposed on observations by the Solar conjunction.  A third and final
paper will map the spectral evolution after emersion from Solar conjunction
(Jan 16, 2022) and up to the re-establishment of pre-outburst appearance,
marking the spectroscopic end of the outburst.  Some quick conclusions can
be drawn from the data presented in this second part of our atlas:\\
(1) the spectra evolution has progressed in a smooth and gradual way,
suggesting that both the expansion of the ejecta through the pre-existing RG
wind and the photo-ionization from the central source did not encountered
sudden discontinuities prior to switch-off of the nuclear burning;\\
(2) the 2021 spectral evolution is similar to that of the 2006 eruption (the
latter much less accurately mapped), with a mirror-appearance of the
triple-peaked line profiles (blue peak brighter than red in 2006, the
opposite in 2021) probably related to the two outbursts occurring at
opposite sides of the binary orbit followed by the red giant and the white
dwarf companion;\\
(3) the evolution of the triple-peaked profiles progressed similarly through
different excitation and ionization degrees, arguing in favor of a
geometrical origin, as in a bipolar outflow nested to an equatorial torus;
a discontinuity occurred around day 30 (at the time when the coronal lines were
rising to their maximum), when the rate of shrinking in the velocity
separation of the red and blue peaks dropped from $-$10 to $-1$~km\,s$^{-1}$ per day;\\
(4) the time-behavior of coronal emission lines appears quite smooth,
both in terms of radiated flux as well as evolution of the triple-peaked
profiles, with a plateau at maximum lasting up to about day 87 (counted from
maximum brightness), and the rise/decline times in agreement with
progressive clearing of the ejecta and then cooling of the WD following
switch-off of the nuclear burning;\\
(5) there seems to exist within RS Oph two distinct types of ejecta: (1)
fast moving ones producing Gaussian-like line profiles (primarely [NII]
and [OIII]), that keep expanding at FWHM=1000~km\,s$^{-1}$ with no sign of
ongoing deceleration during the two months leading up to Solar conjunction,
and (2) slower moving ejecta giving rise to the triple-peaked line profiles
(HeI, HeII, coronal lines, and also Balmer lines at later times), for which
the separation in velocity of blue and red peaks is still shrinking at the
time of Solar conjunction when it reaches an average 330~km\,s$^{-1}$;\\
(6) the "symbiotic band" at 6825 \AA, due to Raman scattering of OVI 1032
\AA\ by neutral hydrogen, appears simultaneously with the coronal lines and
evolves in parallel with them in terms of integrated flux: its very presence
indicates that a sizeable part of the wind of the red giant was neutral
while coronal lines were developing.  In symbiotic stars in quiescence the
6825~\AA\ band is normally observed only in systems undergoing stable
nuclear burning, a condition that appears holding for RS Oph up to around
day +87.}}
\end{center}
\vskip 0.3 cm
\baselineskip 13pt

\noindent
1. \underline{\sc INTRODUCTION}
\bigskip

This is the second of a three-parts atlas of the spectral evolution of the
symbiotic and recurrent nova RS Oph during the 2021 outburst.  The first
part (Munari and Valisa 2021c, hereafter Paper~I) covered the initial phase
of the evolution up to emersion of high ionization emission lines,
corresponding to interval 0 to 18 days past optical maximum (Aug 9 to 27,
2021).  The present Paper II goes from day 19 to 102 (Aug 28 to Nov 19,
2021), when a stop was imposed on observations by Solar conjunction.  A
third and final paper will deal with the period going from emersion from
Solar conjunction (we resumed observations of RS Oph in the pre-dawn sky on
January 16, 2022; Munari and Valisa 2022) up to the re-establishment of
pre-outburst brightness (currently dropped below that level as in all
previous outbursts at similar phases) and of pre-outburst spectral
appearance.

The amount of information stored in a $\sim$daily spectral monitoring in
high-resolution/high-SN/absolute-fluxes of the current outburst of RS Oph
is just immense, and will take a long time and a great effort to be (at
least in part) digested.  Nonetheless, there is a certain urgency to present
at least a quick overlook (even if without much discussion) of some basic
characteristics of such spectral evolution, in support of 
investigations at wavelengths other then optical.

A relevant feature of our atlas is in the great homogeneity maintained
through the whole observing campaign: just three and the same spectrographs
on meter-class telescopes (Asiago 1.22m and 1.82m, Varese 0.84m; all located on
the Italian Alps), identical IRAF-based data reduction and calibration procedures
(as described in the book by Zwitter and Munari 2000), same set of
spectrophotometric standards adopted at all three telescopes, etc. 

Following on the style of Paper~I, a series of 3200-8000 \AA\ spectra at
selected dates are presented to document the overall appearance and
evolution at optical wavelengths, and then the focus shifts to selected
emission lines for which the absolutely-fluxed profiles are presented at 42
distinct dates distributed over the day 19-to-102 time interval.  A few key
plots than follow, presenting: a comparison of lines profiles for the 2006
and 2021 outbursts; the shrinking with time of the separation in velocity of
triple-peaked lines; the evolution in the integrated flux of some lines and
of the underlying true continuum away from lines; the behavior with time of
the FWHM of emission lines; and the evolution of the integrated flux of iron
coronal lines [FeX], [FeXI], [FeXIV] and for reference of [FeVII], as well
as the Raman scattering at 6825 \AA\ of OVI 1032 \AA\ by neutral hydrogen.
\bigskip

\noindent
2. \underline{\sc RS OPH}
\bigskip
 
The current outburst has triggered observations of RS Oph over the whole
electromagnetic spectrum.  As noted in Paper~I, preliminary reports have been
presented by Cheung et al.  (2021a,b) and Wagner \& HESS Collaboration
(2021a,b) for $\gamma$-rays; Enoto et al.  (2021a,b), Ferrigno et al. 
(2021), Luna et al.  (2021), Page (2021a), Page et al.  (2021), Rout et al. 
(2021) and Shidatsu et al.  (2021) for the X-rays; Mikolajewska et al. 
(2021), Munari \& Valisa P (2021a,b), Shore et al.  (2021a,b,c), and
Taguchi et al.  (2021a,b) for the optical; Woodward et al.  (2021a) for the
infrared; and Sokolovsky et al.  (2021) and Williams et al.  (2021) for the
radio.  The results of the search for neutrino emission has been given by
Pizzuto et al.  (2021), and on the polarization of optical light by Nikolov
\& Luna (2021).

Following the release of Paper~I, Page (2021b) announced RS Oph entering the
super-soft X-ray emission phase as later confirmed by Pei et al.  (2021) and
Orio et al.  (2021b,c), with description of X-ray grating spectra given by
Orio et al.  (2021a).  Montez et al.  (2021) reported on the detection and
analysis of extended bipolar X-ray emission stemming from the 2006 eruption. 
Detection of very-high-energy gamma rays from RS Oph have been reported and
analyzed in conjunction with Fermi-LAT data by the H.E.S.S.  Collaboration
(2022) and MAGIC Collaboration (2022).  Zamanov et al.  (2021a) noted the
disappearance of flickering and some optical photometry was provided by
Ricra et al.  (2021), while Fajrin et al.  (2021) reported about their
spectroscopic monitoring in H$\alpha$ as also done by Zamanov et al.  (2021b).
Emersion of coronal emission lines in optical spectra was reported by
Munari et al. (2021), followed by a similar detection at infrared wavelengths
by Woodward et al. (2021). Finally, the photometric and spectroscopic status
of RS Oph after emersion in mid January 2022 from Solar conjunction has been
described by Munari et al. (2022).

\bigskip

\noindent
3. \underline{\sc OBSERVATIONS}
\bigskip

We have used the same telescopes/spectrographs and data acquisition/reduction 
procedures as in Paper~I. 

Echelle observations have been recorded with (1) the 1.82m telescope + REOSC
Echelle spectrograph operated in Asiago by the National Institute of
Astrophysics INAF, that covers in 32 orders and at 22,000 resolving power
(for the usual 1.8-arcsec slit width) the wavelength range 3500-7350~\AA,
without inter-order gaps; and (2) the Varese 0.84m telescope operated by ANS
Collaboration and equipped with an Astrolight Inst.  mk.III Multi-Mode
spectrograph, that in the Echelle configuration covers the 4250-8900 \AA\
range at a 18,000 resolving power (for a 2-arcsec slit) and without
inter-order gaps.

Low resolution observations have been acquired with the B\&C spectrograph on
the 1.22m telescope operated in Asiago by the University of Padova. It
offers a great sensitivity at near-UV wavelengths, and spectra obtained with
the 300 ln/mm grating blazed at 5000 \AA\ allows to cover the 3200-8000~\AA\
range at 2.3 \AA/pix dispersion. The observations here presented at been
recorded at FWHM(PSF)=2~pix sampling.

Data reduction at all three telescopes has been performed in IRAF and has
included all usual steps of correction for bias, dark and flat, sky
subtraction, wavelength calibration, and heliocentric correction. 
Spectrophotometric standards, located close on the sky to RS Oph, have been
observed each night soon before and/or after the RS Oph to achieve an
optimal flux calibration, which was also essential to accurately merge the
individual 30 Echelle orders into a single 1D fluxed spectrum without the
usual dents at the points of junction.

Journals of observations for Echelle and B\&C spectra are provided in
Tables~1 and 2, where the elapsed time $t-t_{\rm max}$ is computed from
maximum optical brightness taken from Paper~1 to have occurred on 
\begin{equation}
t^{V}_{\rm max} = 2459436.18~{\rm JD}  ~~~~~~ 2021 ~{\rm Aug}~ 09.68   ~~~~(\pm 0.05)
\end{equation}
\bigskip

\clearpage

	\begin{table}
	\centering
	\footnotesize
	\caption{Journal of Echelle spectroscopic observations obtained with the
	Varese 0.84m and Asiago 1.82m telescopes.}
	\begin{tabular}{ccrrcc|cccrrc}
        &&\\
	\hline
	&&&&&&&&&&&\\
	 date       & HJD          & t-t$_{\rm max}$& expt & tel.  &&&  date       & HJD          & t-t$_{\rm max}$& expt & tel. \\
	            &              &        &(sec) &       &&&              &              &        &(sec) &     \\
	&&&&&&&&&&&\\
	\hline
	&&&&&&&&&&&\\
	 2021-08-09 & 2459436.300  &   0.12 & 540  & 0.84m &&&    2021-09-07 & 2459465.312  &  29.13 & 5400 & 0.84m \\
	 2021-08-09 & 2459436.337  &   0.16 & 540  & 0.84m &&&    2021-09-08 & 2459466.281  &  30.10 & 3000 & 0.84m \\
	 2021-08-09 & 2459436.373  &   0.19 & 540  & 0.84m &&&    2021-09-09 & 2459467.322  &  31.14 & 2700 & 0.84m \\
	 2021-08-09 & 2459436.404  &   0.22 & 720  & 0.84m &&&    2021-09-10 & 2459468.308  &  32.13 & 4200 & 0.84m \\
	 2021-08-10 & 2459437.320  &   1.14 & 720  & 0.84m &&&    2021-09-11 & 2459469.285  &  33.11 & 2430 & 0.84m \\
	 2021-08-10 & 2459437.389  &   1.21 & 720  & 0.84m &&&    2021-09-12 & 2459470.268  &  34.09 & 2700 & 0.84m \\
	 2021-08-11 & 2459438.342  &   2.16 & 1440 & 0.84m &&&    2021-09-13 & 2459471.271  &  35.09 & 3000 & 0.84m \\
	 2021-08-12 & 2459439.332  &   3.15 & 1200 & 0.84m &&&    2021-09-14 & 2459472.290  &  36.11 & 3600 & 0.84m \\
	 2021-08-13 & 2459440.308  &   4.13 & 900  & 0.84m &&&    2021-09-17 & 2459475.277  &  39.10 & 4200 & 0.84m \\
	 2021-08-14 & 2459441.367  &   5.19 & 1440 & 0.84m &&&    2021-09-21 & 2459479.297  &  43.12 & 3000 & 0.84m \\
	 2021-08-15 & 2459442.310  &   6.13 & 1200 & 0.84m &&&    2021-09-24 & 2459482.269  &  46.09 & 3000 & 0.84m \\
	 2021-08-16 & 2459443.324  &   7.14 & 3000 & 0.84m &&&    2021-09-27 & 2459485.282  &  49.10 & 4050 & 0.84m \\
	 2021-08-17 & 2459444.306  &   8.13 & 900  & 0.84m &&&    2021-09-29 & 2459487.246  &  51.07 & 4050 & 0.84m \\
	 2021-08-18 & 2459445.310  &   9.13 & 1500 & 0.84m &&&    2021-10-01 & 2459489.242  &  53.06 & 4050 & 0.84m \\  
	 2021-08-19 & 2459446.299  &  10.12 & 240  & 1.82m &&&    2021-10-06 & 2459494.240  &  58.06 & 4800 & 0.84m \\
	 2021-08-19 & 2459446.353  &  10.17 & 1500 & 0.84m &&&    2021-10-08 & 2459496.235  &  60.05 & 4800 & 0.84m \\
	 2021-08-20 & 2459447.308  &  11.13 & 1500 & 0.84m &&&    2021-10-10 & 2459498.230  &  62.05 & 3150 & 0.84m \\
	 2021-08-21 & 2459448.301  &  12.12 & 1500 & 0.84m &&&    2021-10-12 & 2459500.231  &  64.05 & 4800 & 0.84m \\
	 2021-08-22 & 2459449.296  &  13.12 & 1500 & 0.84m &&&    2021-10-15 & 2459503.235  &  67.06 & 3600 & 0.84m \\
	 2021-08-22 & 2459449.297  &  13.12 & 540  & 1.82m &&&    2021-10-17 & 2459505.225  &  69.04 & 3600 & 0.84m \\
	 2021-08-23 & 2459450.294  &  14.11 & 2400 & 0.84m &&&    2021-10-19 & 2459507.222  &  71.04 & 4050 & 0.84m \\
	 2021-08-24 & 2459451.314  &  15.13 & 1800 & 0.84m &&&    2021-10-22 & 2459510.217  &  74.03 & 900  & 1.82m \\
	 2021-08-25 & 2459452.296  &  16.12 & 1800 & 0.84m &&&    2021-10-23 & 2459511.218  &  75.04 & 4050 & 0.84m \\
	 2021-08-25 & 2459452.312  &  16.13 & 540  & 1.82m &&&    2021-10-26 & 2459514.223  &  78.04 & 4800 & 0.84m \\
	 2021-08-26 & 2459453.285  &  17.11 & 2400 & 0.84m &&&    2021-10-28 & 2459516.218  &  80.04 & 4050 & 0.84m \\
	 2021-08-27 & 2459454.309  &  18.13 & 600  & 0.84m &&&    2021-11-01 & 2459520.234  &  84.05 & 4800 & 0.84m \\
	 2021-08-28 & 2459455.302  &  19.12 & 2100 & 0.84m &&&    2021-11-04 & 2459523.209  &  87.03 & 4200 & 0.84m \\
	 2021-08-29 & 2459456.292  &  20.11 & 2400 & 0.84m &&&    2021-11-05 & 2459524.210  &  88.03 & 4200 & 0.84m \\
	 2021-08-30 & 2459457.366  &  21.19 & 3600 & 0.84m &&&    2021-11-07 & 2459526.210  &  90.03 & 4500 & 0.84m \\
	 2021-08-31 & 2459458.280  &  22.10 & 2400 & 0.84m &&&    2021-11-11 & 2459530.209  &  94.03 & 3600 & 0.84m \\
	 2021-09-01 & 2459459.287  &  23.11 & 2100 & 0.84m &&&    2021-11-12 & 2459531.209  &  95.03 & 2700 & 0.84m \\
	 2021-09-02 & 2459460.274  &  24.09 & 1500 & 0.84m &&&    2021-11-18 & 2459537.214  & 101.03 & 2700 & 0.84m \\
	 2021-09-03 & 2459461.289  &  25.11 & 3000 & 0.84m &&&    2021-11-19 & 2459538.204  & 102.02 & 2700 & 0.84m \\
	 2021-09-06 & 2459464.314  &  28.13 & 4200 & 0.84m &&&   \\
	&&&&&&&&&&&\\
	\hline
	\end{tabular}
	\end{table}

	\begin{table}
	\centering
	\footnotesize
	\caption{Journal of B\&C spectroscopic observations obtained with the Asiago 1.22m telescope.}
	\begin{tabular}{ccrrcc|cccrrc}
	&&\\
	\hline
	&&&&&&&&&&&\\
	 date       & HJD          & t-t$_{\rm max}$& expt & tel.  &&&  date       & HJD          & t-t$_{\rm max}$& expt & tel. \\
	            &              &        &(sec) &       &&&              &              &        &(sec) &     \\
	\hline
	&&&&&&&&&&&\\
	 2021-08-09 & 2459436.284  &  0.10  & 100 & 1.22m &&&   2021-08-19 & 2459446.281  & 10.10  & 110 & 1.22m   \\
	 2021-08-09 & 2459436.331  &  0.15  & 150 & 1.22m &&&   2021-08-19 & 2459446.330  & 10.15  &  90 & 1.22m   \\
	 2021-08-10 & 2459437.300  &  1.12  & 120 & 1.22m &&&   2021-08-20 & 2459447.307  & 11.13  & 120 & 1.22m   \\
	 2021-08-10 & 2459437.383  &  1.20  & 150 & 1.22m &&&   2021-08-20 & 2459447.369  & 11.19  &  90 & 1.22m   \\
	 2021-08-11 & 2459438.298  &  2.12  &  90 & 1.22m &&&   2021-08-22 & 2459449.321  & 13.14  & 150 & 1.22m   \\
	 2021-08-11 & 2459438.419  &  2.24  & 130 & 1.22m &&&   2021-08-23 & 2459450.279  & 14.10  & 150 & 1.22m   \\
	 2021-08-12 & 2459439.281  &  3.10  &  90 & 1.22m &&&   2021-08-31 & 2459458.297  & 22.12  &  95 & 1.22m   \\
	 2021-08-12 & 2459439.295  &  3.12  & 170 & 1.22m &&&   2021-09-01 & 2459459.282  & 23.10  &  85 & 1.22m   \\
	 2021-08-12 & 2459439.365  &  3.19  & 120 & 1.22m &&&   2021-09-02 & 2459460.267  & 24.09  & 135 & 1.22m   \\
	 2021-08-13 & 2459440.319  &  4.14  & 120 & 1.22m &&&   2021-09-02 & 2459460.354  & 24.17  &  90 & 1.22m   \\
	 2021-08-13 & 2459440.378  &  4.20  & 120 & 1.22m &&&   2021-09-03 & 2459460.349  & 24.17  &  90 & 1.22m   \\
	 2021-08-14 & 2459441.351  &  5.17  & 100 & 1.22m &&&   2021-09-05 & 2459463.267  & 27.09  & 110 & 1.22m   \\
	 2021-08-15 & 2459442.342  &  6.16  & 120 & 1.22m &&&   2021-09-12 & 2459470.284  & 34.10  & 130 & 1.22m   \\
	 2021-08-15 & 2459442.386  &  6.21  &  90 & 1.22m &&&   2021-09-27 & 2459485.264  & 49.08  & 120 & 1.22m   \\
	 2021-08-17 & 2459444.301  &  8.12  & 120 & 1.22m &&&   2021-10-16 & 2459504.233  & 68.05  & 900 & 1.22m   \\
	 2021-08-18 & 2459445.309  &  9.13  & 120 & 1.22m &&&   2021-11-11 & 2459530.201  & 94.02  &1200 & 1.22m   \\
	&&&&&&&&&&&\\
	\hline
	\end{tabular}
	\end{table}

\clearpage
\noindent
4. \underline{\sc OVERALL SPECTRAL APPEARANCE}
\bigskip

The low-resolution spectra obtained with the Asiago 1.22m + B\&C telescope
allow to grasp an overall picture of the evolution of RS Oph during the
19 to 102 day interval under consideration in this Paper~II. A selection of
them is presented in Figure~1.

The brightness of the underlying continuum and the integrated flux of
emission lines have both declined steadily, the latter at a lower pace so
that - as in most novae - the contrast of emission lines appear to increase
with time, expecialy for the nebular ones like [OIII], [OII], [OI], [NeIII],
[NII], [SII], etc. Coronal emission lines started to emerge between
days 25-30, and rapidly increased in intensity with [FeX] remaining always
the strongest of all.  The increase in ionization is well illustrated
by the steady progress with time of the [NeV] 3427 / OIII 3440 ratio. 

It is evident in Figure~1 how the region around 4770~\AA\ has remained
rather free from the presence of emission lines, and consequently it offers
the possibility to evaluate there the flux level of the background
continuum, at blue wavelengths where the direct emission from the stable 
red giant is expected to emerge at later times than in the red.

\begin{figure}[!h]
\centering
\includegraphics[angle=270,width=16.5cm]{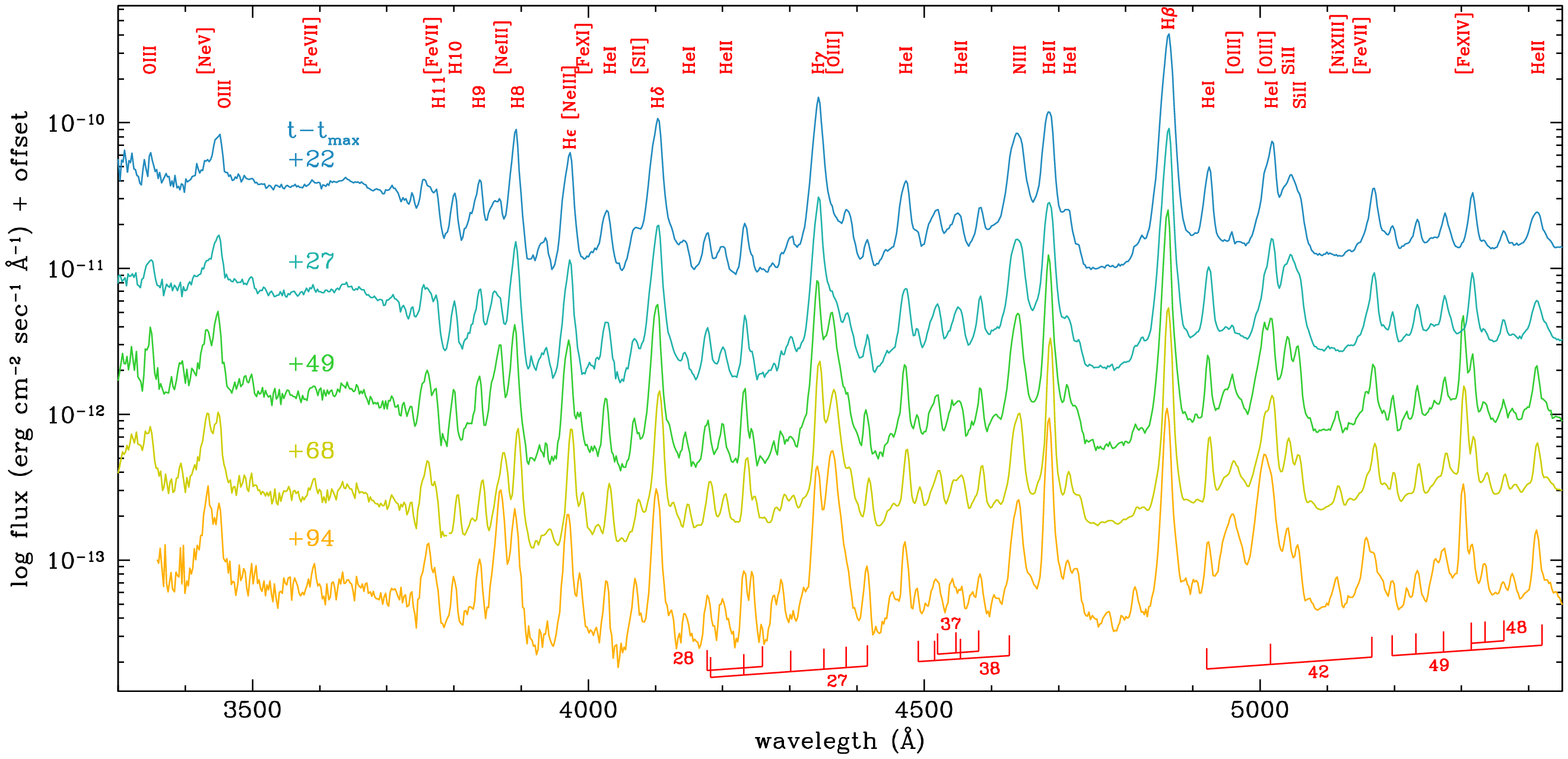}
\includegraphics[angle=270,width=16.5cm]{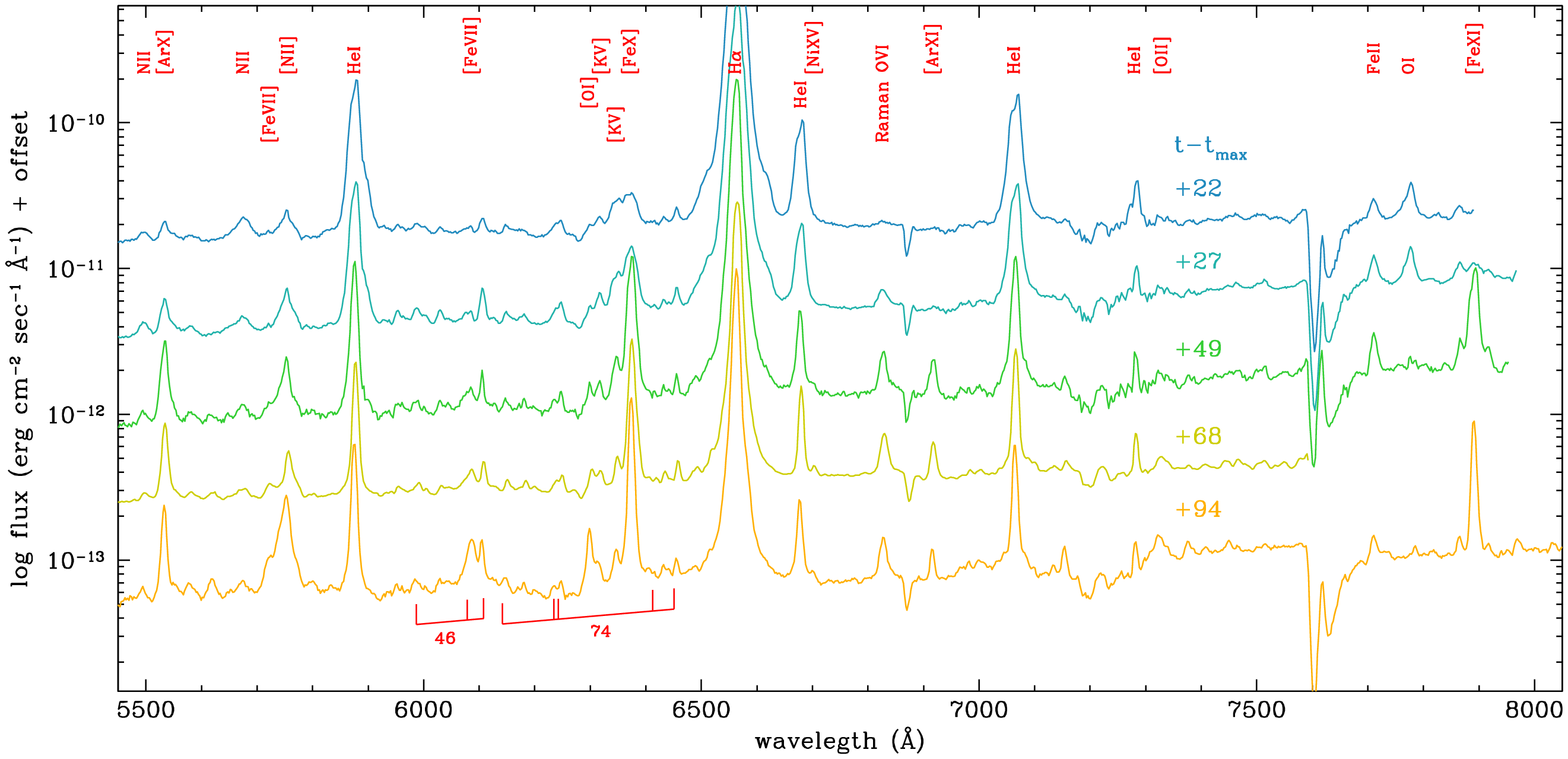}
\caption{Sequence of Asiago 1.22m + B\&C spectra highlighting the evolution
of RS Oph during the 19 to 102 day interval. The most relevant emission
lines are identified at the top, and the comb-markings at the bottom
point to FeII multiplets. The observing epochs are marked as days passed
since maximum brightness $t-t_{\rm max}$ (cf. Table~2).}
\end{figure}

\clearpage
\noindent
5. \underline{\sc COMPARISON OF 2006 AND 2021 LINE PROFILES}
\bigskip

As the photometric evolution of the current outburst is a close replica of
what seen in 2006 and previous events, a marked similarity affects also the
spectra, passing through similar phases at similar times.  There is however
a striking difference: the {\it mirror} appearance of tripled-peaked profiles, as
illustrated in Figure~2.  Throughout the 2006 event and for all lines
presenting a triple-peaked profile, the blue peak has always been brighter
than the red one: the exact contrary is observed for the 2021 outburst,
with the red peak stronger than the blue one at all phases and for all lines
as illustrated by the time-sequences of profiles presented in Figures~6
to 11.  With all probability this difference relates to the fact the 2006
and 2021 outbursts occurred on opposite sides of the binary orbit.

The triple-peaked profile is present or at least recognizable in nearly all
lines observed in 2021, with the exception of  [NII], [OIII] and a few
others. Also Balmer lines evolved to a triple-peaked profile by the time the
Solar conjunction was approaching.  

For all triple-peaked line profiles (cf HeI, HeII, or coronal lines in
Figures~6 to 10), the side peaks (i.e.  the red and the blue peak) decline
faster than the central peak.  At late times, for some of the riple-peaked
lines (eg.  [FeXIV] 5303) only the central peak remained visible.

Balmer lines presented a lower degree of "mirror symmetry" when comparing
2006 and 2021 profiles.  From Figure~2, a triple-peaked profile seems
present in H$\beta$ in 2006 already at day 30, but with a reversed blue/red
peak ratio compared to the other 2006 lines and with a prominent central
peak, while H$\alpha$ looks rather symmetrical.  With a certain degree of
imagination, a triple-peaked profile can be recognized also in the 2021
H$\beta$ profile for day 30 in Figure~2 but, differently from 2006, with a
insignificant central peak and a blue/red peak ratio similar to that of the
other lines and not reversed.

When comparing the relative displacement of the narrow (from the ionized
wind) and very broad (from expanding ejecta) components in the [NII] 5755
profile, a "mirror symmetry" is present comparing 2006 to 2021 profiles in
Figure~2.

\begin{figure}[!h]
\centering
\includegraphics[width=9cm]{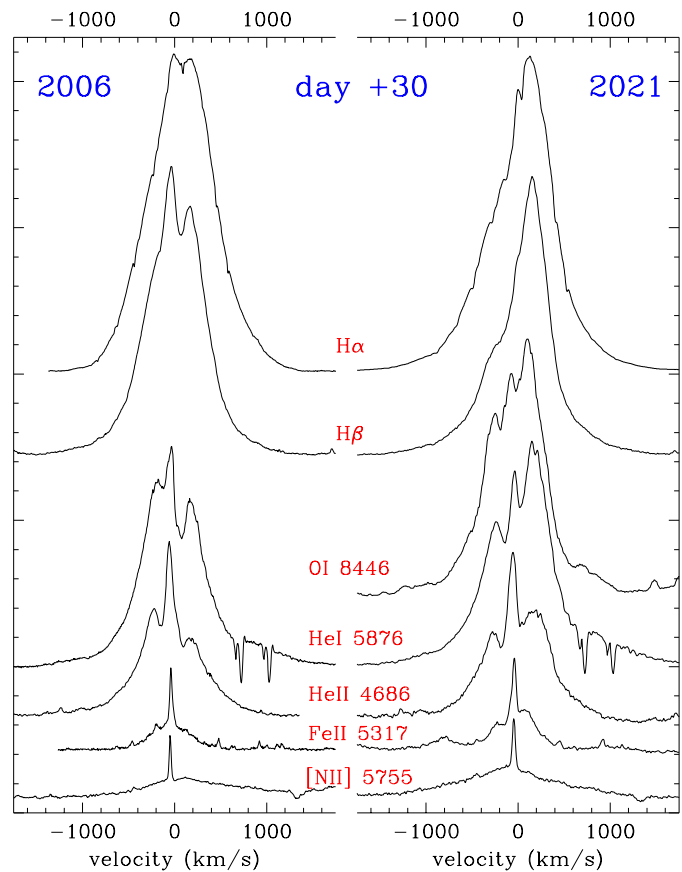}
\caption{A comparison of the profile of a few representative lines, at day
+30 from optical maximum, for the 2006 (from Munari et al. 2007) and 
2021 outbursts.} 
\end{figure}

\clearpage
\noindent
6. \underline{\sc VELOCITY SEPARATION OF TRIPLE-PEAKED LINE PROFILES}
\bigskip

The velocity separation ($\Delta$V) of red and blue peaks has been measured on the 
Echelle spectra of Figures~6 to 11 for a number of emission lines and its
progression with time is plotted in Figure~3.

Higher ionization lines as HeII, [FeX] and [FeXIV] behave all rather similarly, with
a separation $\sim$60~km~s$^{-1}$ larger than for lower excitation lines
like FeII and OI. The HeI lines locate neatly in between, with the green
line in Figure~3 being a spline fitting to all measurable HeI lines, in
particular to 6678 and 5015 singlet lines, and 5876 and 7065 triplet lines.

The behavior of $\Delta$V for HeI lines (copied by HeII) is characterized by
a marked change in slope taking place during the day 30 to 50 interval (this
is the same time interval over which coronal lines reach in turn their
respective plateau maxima, from [FeX] around day 30 to [FeXIV] about day 50;
see Figure sect.  7 and Figure~5 below): namely, before day 30 $\Delta$V
steadily reduces by $-$10 km~s$^{-1}$ each day, while after day 50 such
shrinking is drastically slowed to a mere $-$1 km~s$^{-1}$ per day.

\begin{figure}[!h]
\centering
\includegraphics[width=12.6cm]{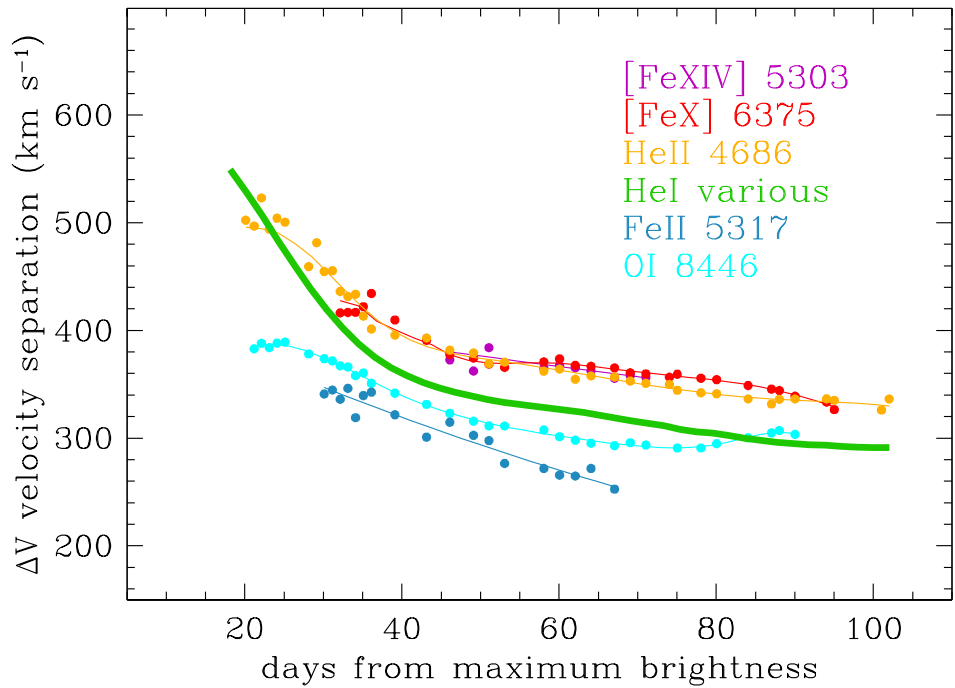}
\caption{Velocity separation of the blue and red peaks of the triple-peaked 
profiles for selected lines on the 2021 Echelle spectra of RS Oph presented
in Figures~6 to 11. By "HeI various" we mean a spline fitting to all measurable 
HeI lines, in particular the 6678 and 5015 singlet lines, and 5876 and 7065 triplet
lines.}
\end{figure}
\bigskip
\bigskip

\noindent
7. \underline{\sc EVOLUTION OF INTEGRATED FLUX AND WIDTH OF BALMER LINES}
\bigskip

The evolution in terms of integrated flux of H$\alpha$ and H$\beta$ emission
lines is presented in the top panel of Figure~4, compared to the background
brightness measured at 4770~\AA.  The decline in flux of H$\alpha$ and
H$\beta$ has been very smooth and very similar, with measurement of initial
H$\alpha$ values being probably affected by P-Cyg absorptions, overlapping
nearby lines and the blackground glare of the flash-ionized wind of the
red-giant.  A marked acceleration of the decline in flux sets around day
+85, contemporaneous with the coronal lines abandoning the plateau maximum
and beginning their rapid descent (cf.  next section and Figure~5).  The
change in slope of H$\alpha$ and H$\beta$ has been somewhat mirrored by a
similar change visible in the intensity of the background measured at
4770~\AA.  At early times the flux in the continuum declines must faster
that flux in the Balmer lines, and only after day 30 they behave in
parallel.  The fluxes for 4770~\AA\ and H$\beta$ are tabulated in Table~3.

The evolution of the width of H$\alpha$ and H$\beta$ emission lines is
illustrated in the bottom panel of Figure~4.  Both behave very smoothly and
rather similarly, with differences at latest epochs being probably spurious
and induced by the appearance of the triple-peaked profile occuring first in
H$\gamma$, then H$\beta$, and only last in H$\alpha$. 

From the bottom panel of Figure~4 it is evident how the width of all lines
displaying, sooner or later, a triple-peaked profile ended converging toward
FWHM$\approx$330~km\,s$^{-1}$, a value than was still shrinking at the time
when Solar conjunction put a halt to the observations.  On the contrary,
emission lines like [NII] and [OIII] showing a profile characterized by a
single broad Gaussian (ignoring the superimposed sharp peak originating in
the kinematically quiet and flashed wind of the red giant), had already
reached their terminal width at much earlier times (day $\sim$45), with no
further significantive shrinking afterward.  Their terminal width,
FWHM$\approx$1000~km\,s$^{-1}$, is three times broader than for
triple-peaked profiles.

It therefore seems that within RS Oph co-existed two distinct types of ejecta:
($i$) a faster moving one, giving rise to the Gaussian-like broad profiles for [NII]
and [OIII], leaving the central binary in a sort of symmetrical arrangement at
a terminal $\sim$425~km\,s$^{-1}$ velocity (taking it equal to
FWHM/$2 \sqrt{2\,\ln 2}$); and ($ii$) a slower moving one, giving rise to the
triple-peaked profiles, with the material confined in the bipolar regions
moving at $\sim$$|$165$|$~km\,s$^{-1}$ away from the material that originates the
central peak and which is at rest with the binary system.

\begin{figure}[!h]
\centering
\includegraphics[width=11.8cm]{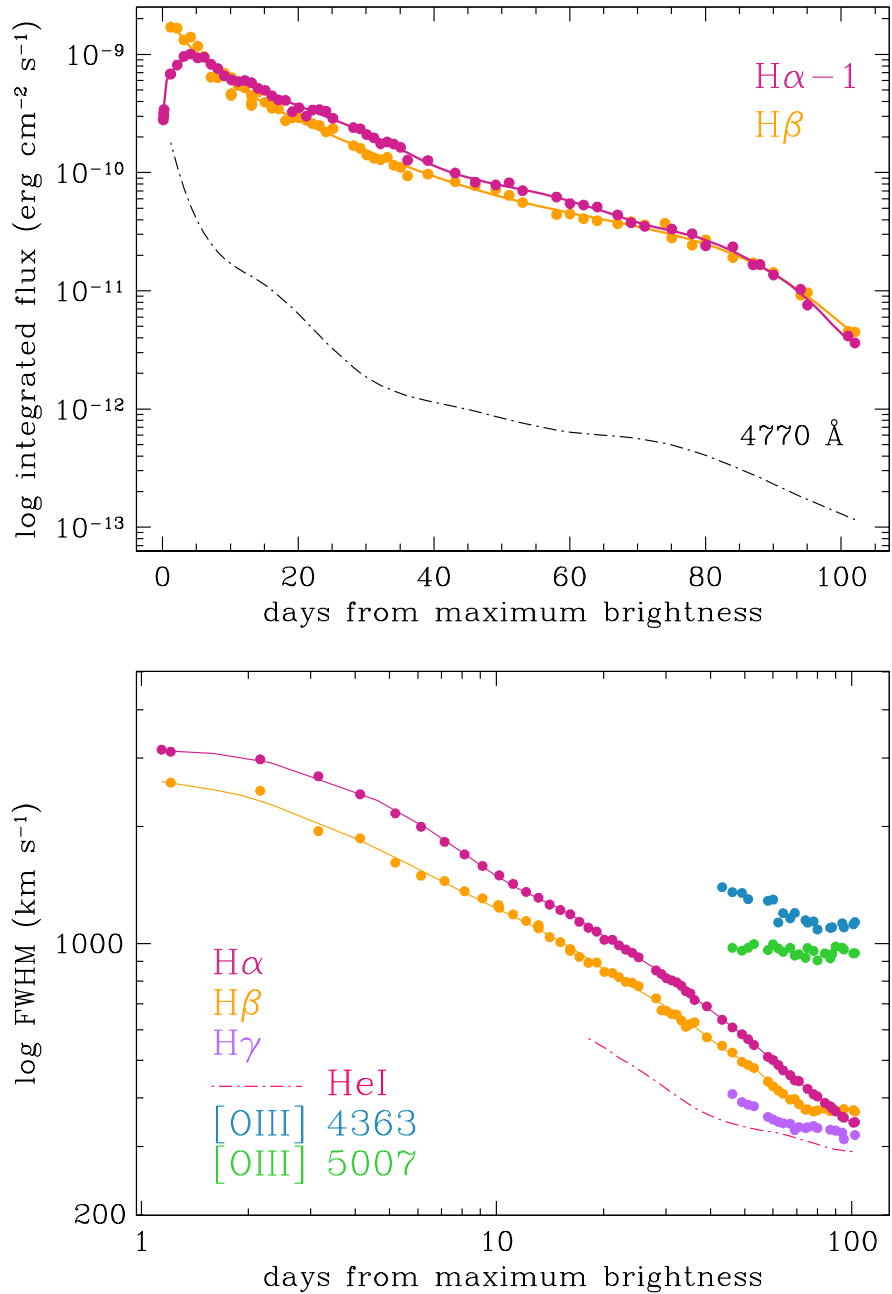}
\caption{{\em Top panel}: evolution of the integrated flux of H$\alpha$ and
H$\beta$ emission lines (log scale), measured on the 2021 Echelle spectra of
RS Oph presented in Figures~6 to 11.  Note the "$-$1" offset for H$\alpha$ in
the top panel, for an easier comparison with H$\beta$.  The
dot-dashed line is the flux in the background continuum measured at 4770
\AA\ (and expressed in erg cm$^{-2}$ s$^{-1}$ \AA$^{-1}$).  {\em Bottom panel}:
evolution of the FWHM of the H$\alpha$ and H$\beta$ emission lines (log-log
plot), and for comparison also of other lines. The
dot-dashed line is the "HeI various" from previous Figure~3 (cf. sect.5).}
\end{figure}

\clearpage
\noindent
8. \underline{\sc EVOLUTION OF CORONAL AND OVI EMISSION LINES}
\bigskip

A hallmark of the 2021 and all previous outbursts of RS Oph is the
development of strong coronal lines.  We have measured the integrated flux
of the iron coronal lines in the Echelle spectra of Figures~6 and 7, and
tabulated them in Table~3 with a graphical presentation given in Figure~5. 
In the same figure, we have plotted for comparison the behavior of the
continuum at 4770~\AA\ and of the Raman scattering by neutral hydrogen at
6825~\AA\ of OVI~1032.

The coronal lines started to appear with day 23, when the optical thickness
of the expanding ejecta lowered sufficiently to allow ionizing photon to
spread through.  After a rapid rise in intensity, they all levelled off at a
plateau maximum which was very smooth and that mildly declined in pace with
the continuum background emission.  Even if their appearance and their
reaching maximum brightness happened in a succession paralleling their
ionization degree, all the coronal lines left the plateau maximum at
basically the same time, around day 87, to begin a rapid decline consistent
with a switching-off of the nuclear burning on the white dwarf of RS Oph.

Of special interest is the presence and behavior of the "symbiotic band" at
6825~\AA.  This band is only observed in the symbiotic stars that in quiescence
are powered by nuclear burning of the material accreted by their WDs.  For a
long time its origin has remained unidentified, until the mystery was solved
by Schmid (1989) is terms of Raman scattering at 6825 and 7088~\AA\ of the
ultraviolet resonance doublet OVI~1032, 1038~\AA\ by neutral hydrogen present within
the binary.  During the quiescence in between the outbursts, RS Oph does not
burn on its WD and does not display the 6825 and 7088~\AA\ bands (Munari and
Zwitter 2002), which are observable only during nova outbursts (the 7088~\AA\
band is much weaker than 6825~\AA\ and essentially lost in the bright wings
of HeI 7065).  The behavior of 6825~\AA\ in Figure~5 is rather similar to
the coronal lines in terms of time of appearance, rising rate, plateau
maximum, and onset of the decline.  The very presence of 6825~\AA\ proves
that a sizeable fraction of the wind of the red giant remained neutral
during the development of the coronal lines, and reinforces the notion that 
nuclear burning was probably active through all that period. 

\begin{figure}[!h]
\centering
\includegraphics[width=14.3cm]{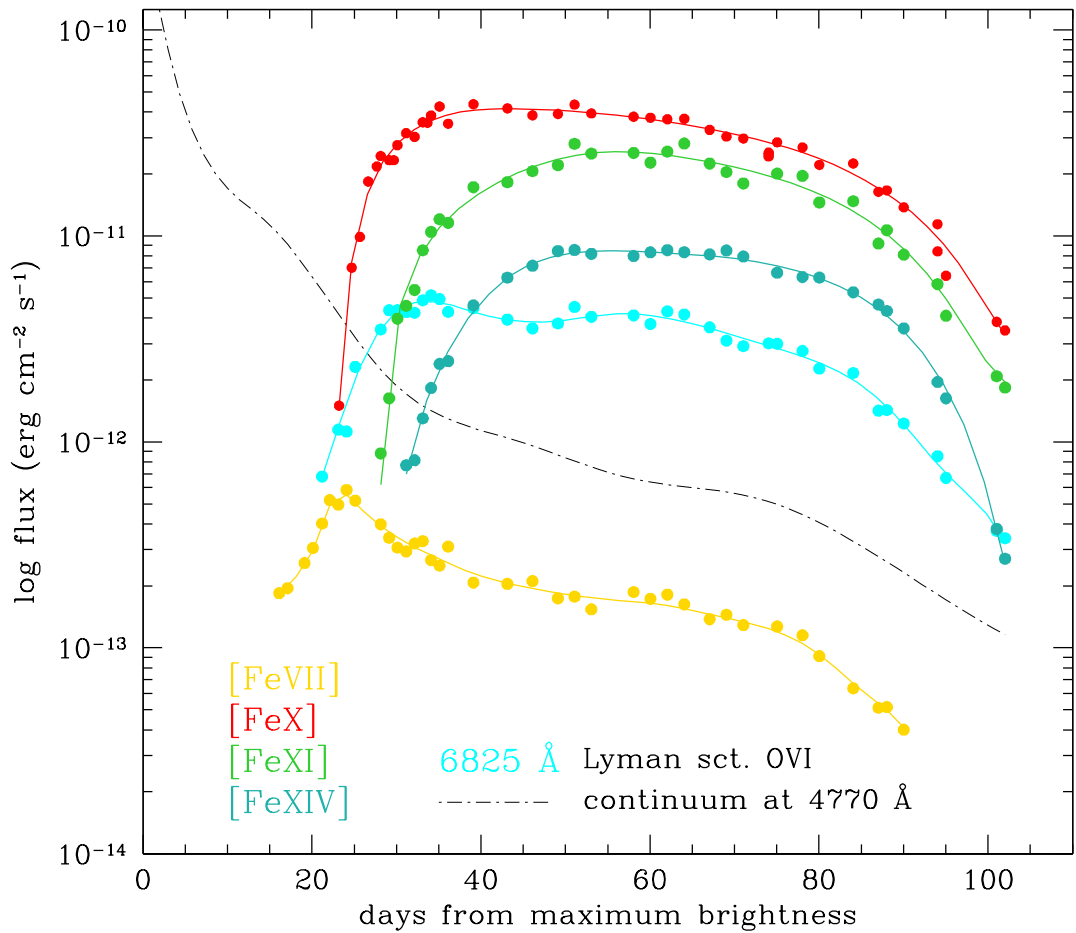}
\caption{Evolution along the 2021 outburst of the integrated flux of the
coronal lines [FeX] 6375, [FeXI] 7892, and [FeXIV] 5303 \AA, and for
comparison of [FeVII] 6087 \AA, the "symbiotic band" at 6825~\AA\ (Raman
scattering by neutral hydrogen of OVI~1032), and of the continuum at
4770~\AA\ (the latter expressed in erg cm$^{-2}$ s$^{-1}$ \AA$^{-1}$).}
\end{figure}

	\begin{table}
	\caption{Logarithm of the flux in the continuum at 4770~\AA\ (in erg cm$^{-2}$
        s$^{-1}$ \AA$^{-1}$) and the logarithm of the integrated flux (in erg
        cm$^{-2}$ s$^{-1}$) of H$\beta$, [FeVII] 6087, [FeX] 6375, [FeXI]
        7892, and [FeXIV] 5303 emission lines from the Echelle spectra
        listed in Table~1 and displayed in Figure~6 to 11.}
	\centering
	\footnotesize
	\begin{tabular}{rrrrrrr}
        &&\\
	\hline
	&&\\
	\multicolumn{1}{r}{t-t$_{\rm max}$}&
	\multicolumn{1}{r}{4770 \AA}&
	\multicolumn{1}{r}{H$\beta$~~}&
	\multicolumn{1}{c}{[FeVII]}&
	\multicolumn{1}{c}{[FeX]}&
	\multicolumn{1}{c}{[FeXI]}&
	\multicolumn{1}{c}{[FeXIV]} \\
	&&\\
	\hline
	&&\\
	   1.14   &    $-$9.731    &   $-$8.756    &                &             &                  &                 \\
	   1.21   &    $-$9.747    &   $-$8.770    &                &             &                  &                 \\
	   2.16   &    $-$9.951    &   $-$8.778    &                &             &                  &                 \\
	   3.15   &   $-$10.129    &   $-$8.878    &                &             &                  &                 \\
	   4.13   &   $-$10.280    &   $-$8.853    &                &             &                  &                 \\
	   5.19   &   $-$10.415    &   $-$8.930    &                &             &                  &                 \\
	   6.13   &   $-$10.514    &   $-$9.017    &                &             &                  &                 \\
	   7.14   &   $-$10.602    &   $-$9.193    &                &             &                  &                 \\
	   8.13   &   $-$10.671    &   $-$9.197    &                &             &                  &                 \\
	   9.13   &   $-$10.728    &   $-$9.160    &                &             &                  &                 \\
	  10.12   &   $-$10.773    &   $-$9.340    &                &             &                  &                 \\
	  10.17   &   $-$10.775    &   $-$9.196    &                &             &                  &                 \\
	  11.13   &   $-$10.812    &   $-$9.264    &                &             &                  &                 \\
	  12.12   &   $-$10.846    &   $-$9.283    &                &             &                  &                 \\
	  13.12   &   $-$10.879    &   $-$9.402    &                &             &                  &                 \\
	  14.11   &   $-$10.912    &   $-$9.318    &                &             &                  &                 \\
	  15.13   &   $-$10.952    &   $-$9.405    &                &             &                  &                 \\
	  16.13   &   $-$10.995    &   $-$9.443    &   $-$12.734    &             &                  &                 \\
	  17.11   &   $-$11.041    &   $-$9.461    &   $-$12.710    &             &                  &                 \\
	  19.12   &   $-$11.147    &   $-$9.537    &   $-$12.588    &             &                  &                 \\
	  20.11   &   $-$11.203    &   $-$9.535    &   $-$12.515    &             &                  &                 \\
	  21.19   &   $-$11.265    &   $-$9.556    &   $-$12.396    &             &                  &                 \\
	  22.10   &   $-$11.319    &   $-$9.586    &   $-$12.282    &             &                  &                 \\
	  23.11   &   $-$11.378    &   $-$9.601    &   $-$12.305    &  $-$11.824  &                  &                 \\
	  24.09   &   $-$11.435    &   $-$9.656    &   $-$12.234    &  $-$11.154  &                  &                 \\
	  25.11   &   $-$11.493    &   $-$9.628    &   $-$12.285    &  $-$11.005  &                  &                 \\
	  28.13   &   $-$11.649    &   $-$9.772    &   $-$12.400    &  $-$10.612  &    $-$12.056     &                 \\
	  29.13   &   $-$11.693    &   $-$9.795    &   $-$12.465    &  $-$10.632  &    $-$11.789     &                 \\
	  30.10   &   $-$11.731    &   $-$9.850    &   $-$12.513    &  $-$10.559  &    $-$11.402     &                 \\
	  31.14   &   $-$11.768    &   $-$9.879    &   $-$12.532    &  $-$10.501  &    $-$11.339     & $-$12.113       \\
	  32.13   &   $-$11.798    &   $-$9.891    &   $-$12.493    &  $-$10.519  &    $-$11.262     & $-$12.089       \\
	  33.11   &   $-$11.825    &   $-$9.871    &   $-$12.482    &  $-$10.448  &    $-$11.070     & $-$11.886       \\
	  34.09   &   $-$11.849    &   $-$9.939    &   $-$12.574    &  $-$10.415  &    $-$10.980     & $-$11.738       \\
	  35.09   &   $-$11.870    &   $-$9.956    &   $-$12.601    &  $-$10.372  &    $-$10.919     & $-$11.621       \\
	  36.11   &   $-$11.889    &  $-$10.028    &   $-$12.508    &  $-$10.455  &    $-$10.936     & $-$11.608       \\
	  39.10   &   $-$11.933    &  $-$10.013    &   $-$12.683    &  $-$10.360  &    $-$10.763     & $-$11.337       \\
	  43.12   &   $-$11.981    &  $-$10.078    &   $-$12.690    &  $-$10.380  &    $-$10.739     & $-$11.202       \\
	  46.09   &   $-$12.020    &  $-$10.111    &   $-$12.676    &  $-$10.414  &    $-$10.685     & $-$11.144       \\
	  49.10   &   $-$12.064    &  $-$10.143    &   $-$12.760    &  $-$10.407  &    $-$10.656     & $-$11.074       \\
	  51.07   &   $-$12.093    &  $-$10.193    &   $-$12.751    &  $-$10.362  &    $-$10.553     & $-$11.068       \\
	  53.06   &   $-$12.121    &  $-$10.253    &   $-$12.813    &  $-$10.405  &    $-$10.600     & $-$11.087       \\
	  58.06   &   $-$12.180    &  $-$10.355    &   $-$12.729    &  $-$10.421  &    $-$10.597     & $-$11.097       \\
	  60.05   &   $-$12.196    &  $-$10.351    &   $-$12.762    &  $-$10.426  &    $-$10.644     & $-$11.079       \\
	  62.05   &   $-$12.208    &  $-$10.390    &   $-$12.742    &  $-$10.433  &    $-$10.590     & $-$11.069       \\
	  64.05   &   $-$12.218    &  $-$10.407    &   $-$12.789    &  $-$10.431  &    $-$10.551     & $-$11.079       \\
	  67.06   &   $-$12.232    &  $-$10.433    &   $-$12.861    &  $-$10.485  &    $-$10.648     & $-$11.089       \\
	  69.04   &   $-$12.243    &  $-$10.414    &   $-$12.840    &  $-$10.517  &    $-$10.690     & $-$11.070       \\
	  71.04   &   $-$12.258    &  $-$10.442    &   $-$12.889    &  $-$10.527  &    $-$10.745     & $-$11.100       \\
	  74.04   &   $-$12.290    &  $-$10.429    &   $-$12.906    &  $-$10.601  &    $-$10.711     & $-$11.137       \\
	  75.04   &   $-$12.303    &  $-$10.553    &   $-$12.896    &  $-$10.545  &    $-$10.697     & $-$11.178       \\
	  78.04   &   $-$12.353    &  $-$10.614    &   $-$12.939    &  $-$10.570  &    $-$10.708     & $-$11.199       \\
	  80.04   &   $-$12.392    &  $-$10.570    &   $-$13.040    &  $-$10.655  &    $-$10.837     & $-$11.202       \\
	  84.05   &   $-$12.482    &  $-$10.719    &   $-$13.196    &  $-$10.648  &    $-$10.831     & $-$11.273       \\
	  87.03   &   $-$12.556    &  $-$10.763    &   $-$13.291    &  $-$10.786  &    $-$11.037     & $-$11.333       \\
	  88.03   &   $-$12.582    &  $-$10.779    &   $-$13.288    &  $-$10.779  &    $-$10.972     & $-$11.364       \\
	  90.03   &   $-$12.635    &  $-$10.845    &   $-$13.398    &  $-$10.861  &    $-$11.091     & $-$11.449       \\
	  94.03   &   $-$12.741    &  $-$11.038    &                &  $-$11.074  &    $-$11.234     & $-$11.709       \\
	  95.03   &   $-$12.767    &  $-$11.016    &                &  $-$11.193  &    $-$11.387     & $-$11.788       \\
	 101.03   &   $-$12.913    &  $-$11.342    &                &  $-$11.417  &    $-$11.681     & $-$12.423       \\
	 102.02   &   $-$12.935    &  $-$11.349    &                &  $-$11.459  &    $-$11.736     & $-$12.567       \\
	&&\\
	\hline
	\end{tabular}
	\end{table}

\clearpage
\noindent
9. \underline{\sc PICTORIAL EVOLUTION OF SELECTED EMISSION LINES}
\bigskip

In Figures 6 to 11 we present the temporal evolution of a sample of emission
lines from our Echelle spectra.  The $t-t_{\rm max}$ time from maximum (in
days) is marked in red next to each spectrum.  The spectra are plotted in a
log(flux) + offset scale, which makes visible both faint features on the
wings as well as the vastly brighter core.  Earlier epochs are covered in
Paper~I, from which for continuity and overlap we take the first two spectra
for day 16 and 17.

\begin{figure}[!b]
\centering
\includegraphics[width=15cm]{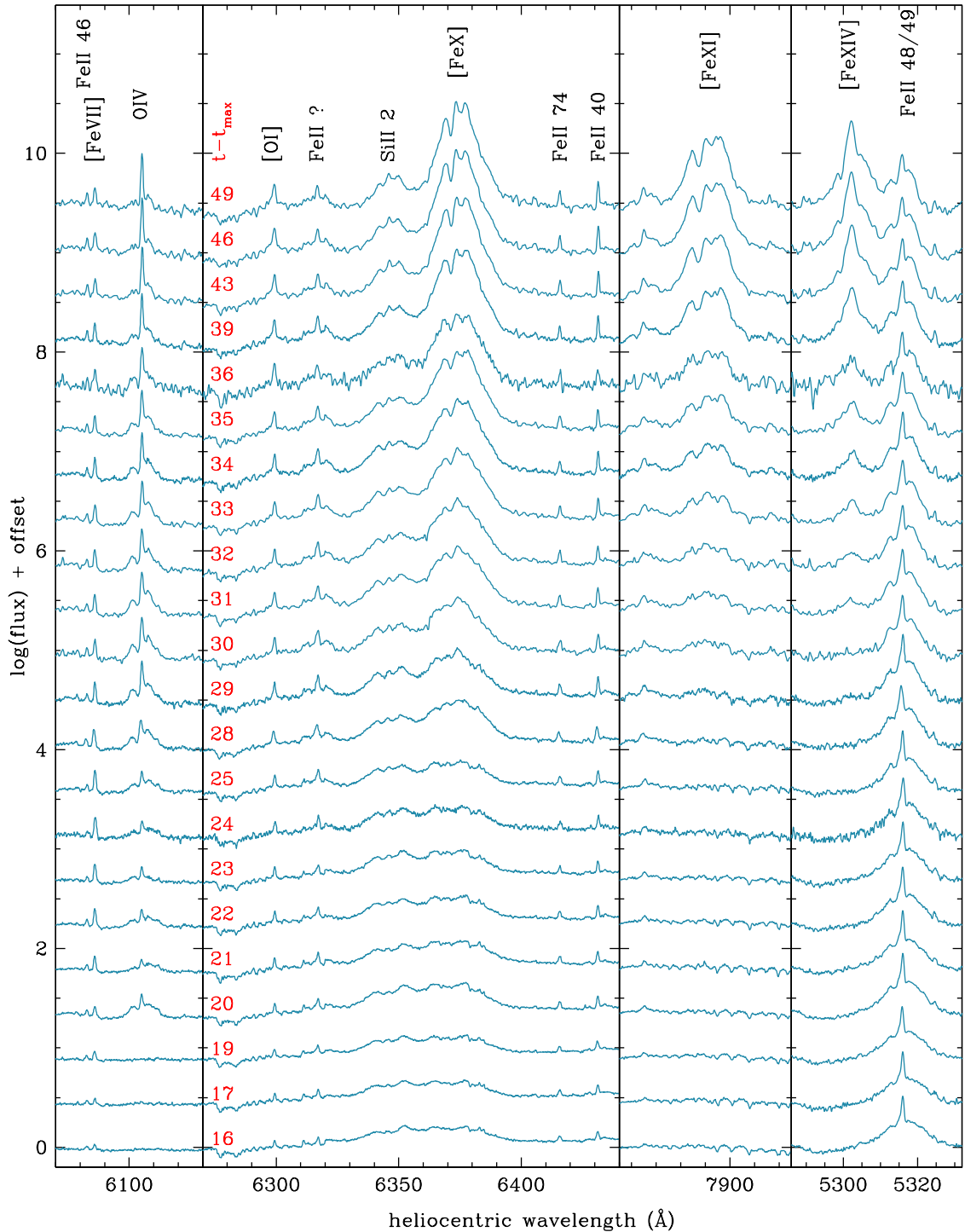}
\caption{Evolution from day 16 to 102 of selected emission lines in our
Echelle spectra of the 2021 outburst of RS Oph (cf Table~1).  The $t-t_{\rm
max}$ time from maximum (in days) is marked in red next to each spectrum.
Identification of OIV 6106 following R.M. Wagner (private comm.).}
\end{figure}

\begin{figure}[!b]
\centering
\includegraphics[width=17cm]{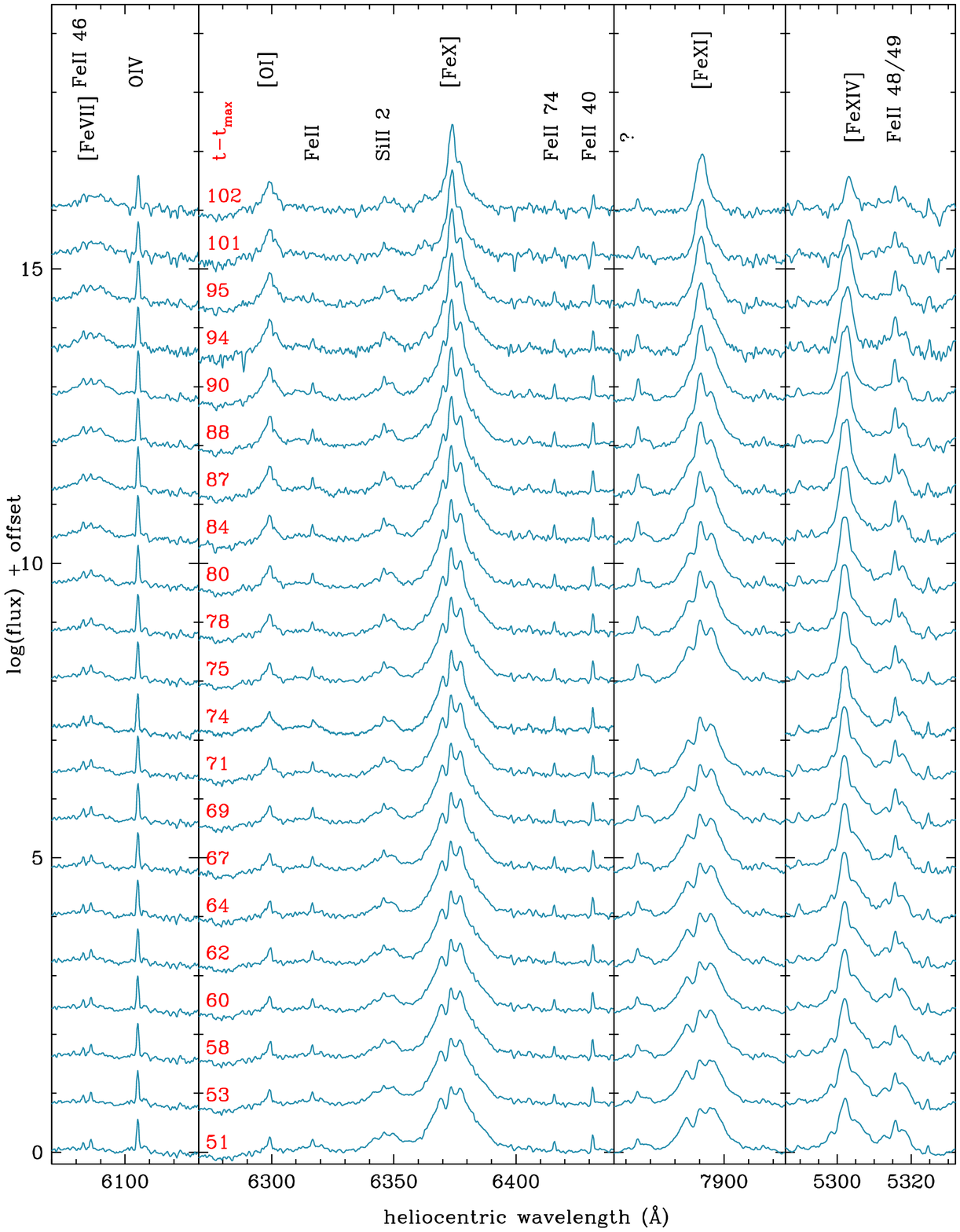}
\caption{continues from Figure~6.}
\end{figure}

\begin{figure}[!b]
\centering
\includegraphics[width=17cm]{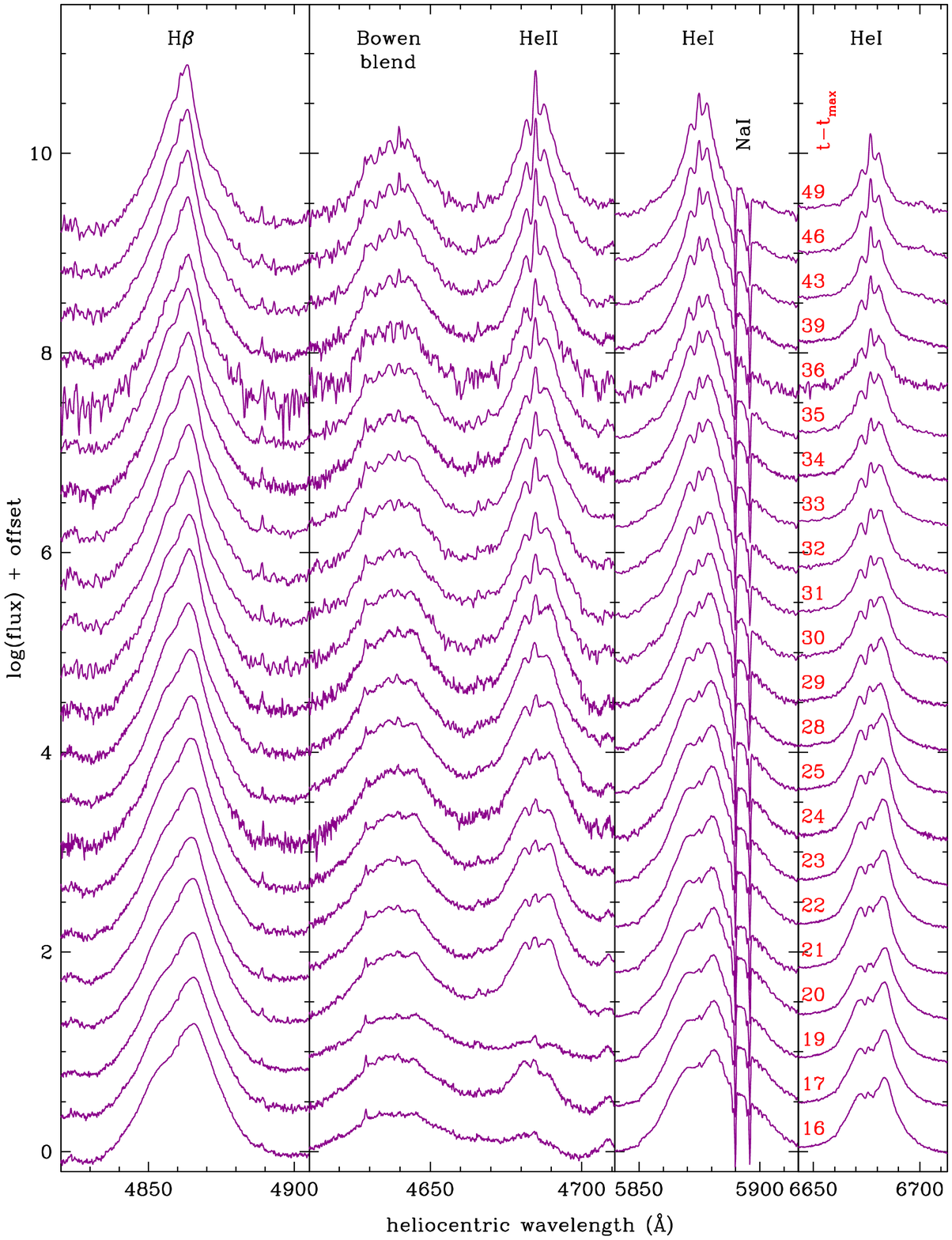}
\caption{continues from Figure~6.}
\end{figure}

\begin{figure}[!b]
\centering
\includegraphics[width=17cm]{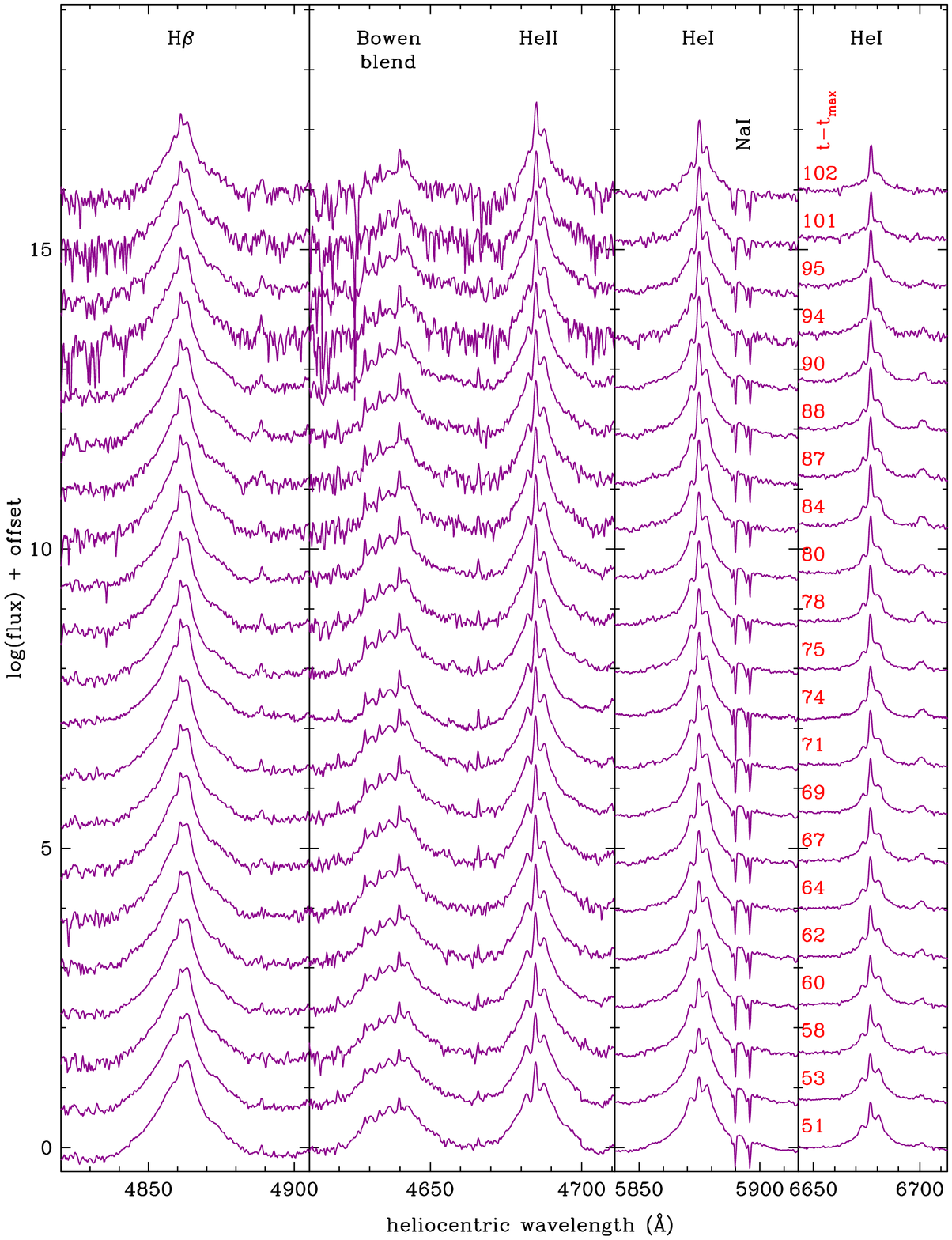}
\caption{continues from Figure~6.}
\end{figure}

\begin{figure}[!b]
\centering
\includegraphics[width=17cm]{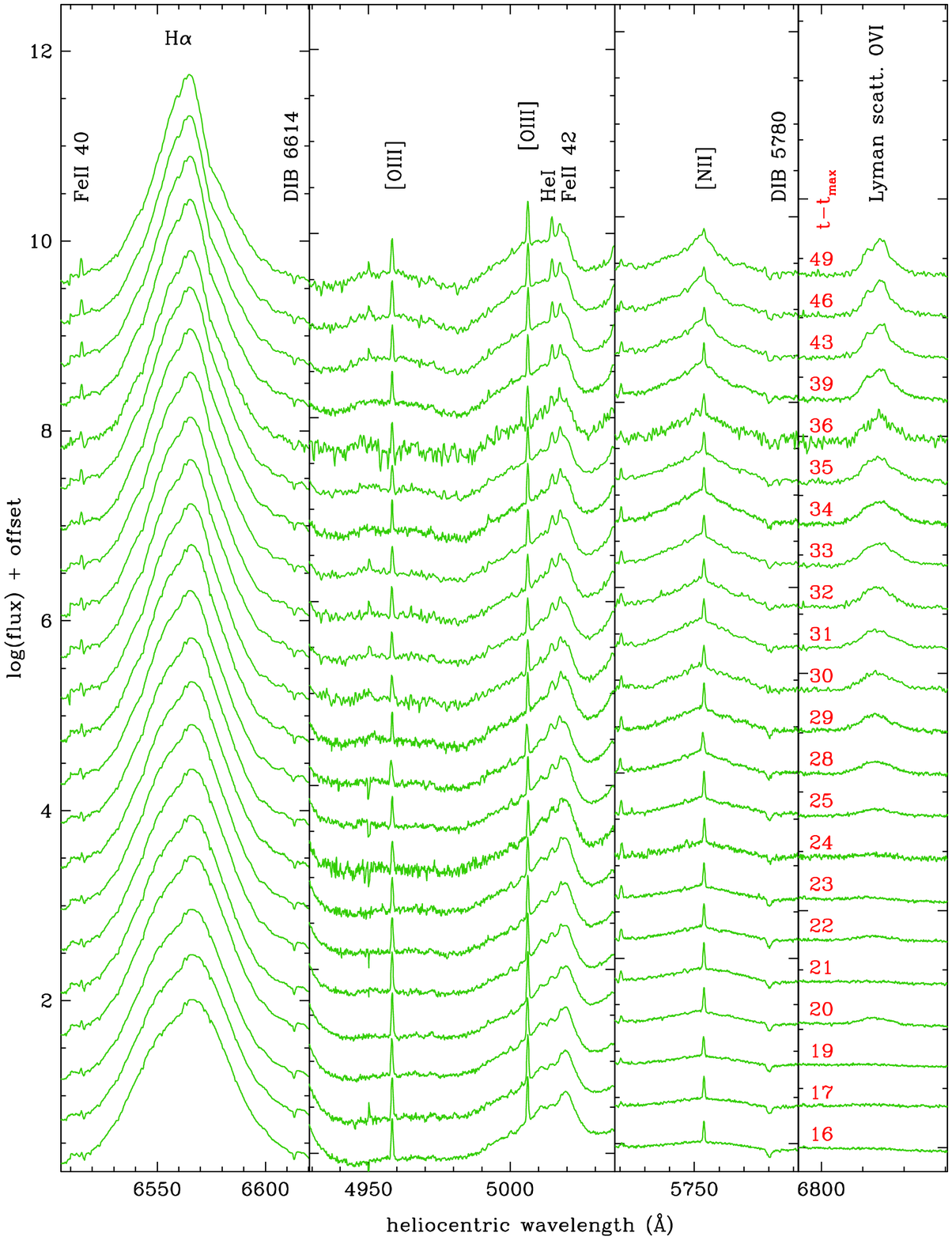}
\caption{continues from Figure~6.}
\end{figure}

\begin{figure}[!b]
\centering
\includegraphics[width=17cm]{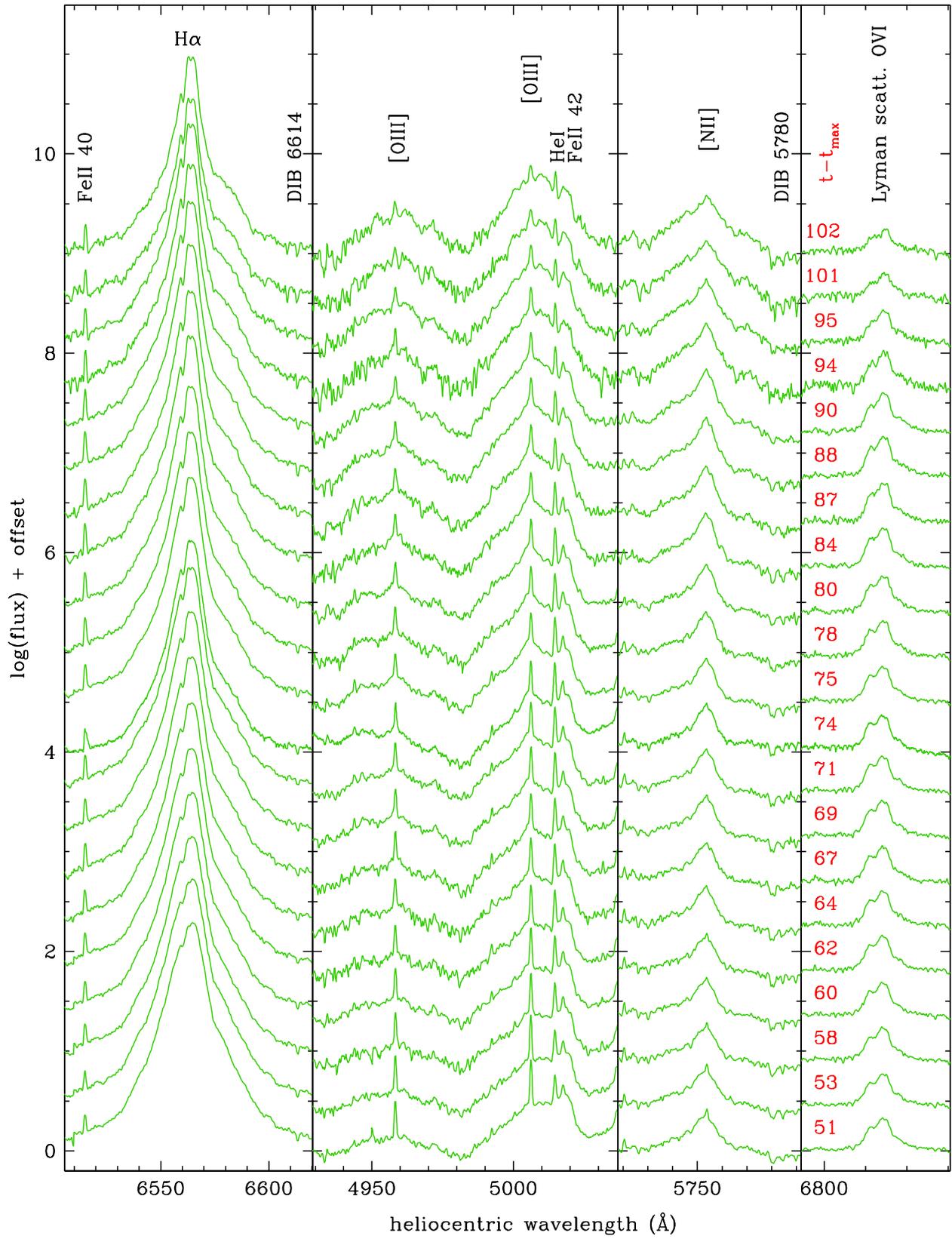}
\caption{continues from Figure~6.}
\end{figure}

\clearpage
10. \underline{\sc REFERENCES}
\bigskip

Cheung C.~C., Ciprini S., Johnson T.~J., 2021a, ATel, 14834

Cheung C.~C., Johnson T.~J., Mereu I., et al., 2021b, ATel, 14845

Enoto T., Maehara H., Orio M., et al., 2021a, ATel, 14850

Enoto T., Orio M., Fabian A., et al., 2021b, ATel, 14864

Evans A., Bode M. F., O'Brien T. J., Darnley M. J., eds., 2008, RS Ophiuchi (2006) and the Recurrent Nova 

{~~~~~}Phenomenon, ASPC, 401  

Fajrin M., Imaduddin I., Malasan H.~L., et al., 2021, ATel, 14909

Ferrigno C., Savchenko V., Bozzo E., et al., 2021, ATel, 14855

H.E.S.S. Collaboration, et al., 2022, arXiv:2202.08201

Luna G.~J.~M., Jimemez-Carrera R., Enoto T., et al., 2021, ATel, 14872

MAGIC Collaboration, et al., 2022, arXiv:2202.07681

Mikolajewska J., Aydi E., Buckley D., et al., 2021, ATel, 14852

Montez R., Luna G.~J.~M., Mukai K., et al. , 2021, arXiv:2110.04315

Munari U., et al., 2007, BaltA 16, 46 

Munari U., Valisa P., 2021a, ATel, 14840

Munari U., Valisa P., 2021b, ATel, 14860

Munari U., Valisa P., 2021c, arXiv 2109.01101 (Paper I)

Munari U., Valisa P., Ochner P., 2021, ATel, 14895

Munari U., Valisa P., Dallaporta S., 2022, ATel, 15169

Munari U., Zwitter T., 2002, A\&A 383, 188

Nikolov Y., Luna G.~J.~M., 2021, ATel, 14863

Orio M., Behar E., Drake J., et al. , 2021a, ATel, 14906

Orio M., Gendreau K., Pei S., et al., 2021b, ATel, 14926

Orio M., Gendreau K., Pei S., et al., 2021c, ATel, 14954

Page, K.~L. 2021a, ATel, 14885

Page K.~L., 2021b, ATel, 14894

Page K.~L., Osborne J.~P., Aydi E., 2021, ATel, 14848

Pei S., Orio M., Gendreau K., et al., 2021, ATel, 14901

Pizzuto A., Vandenbroucke J., Santander M., IceCube Collaboration, 2021, ATel, 14851

Ricra J., Vannini J., Baella N.~O., 2021, ATel, 14972

Rout S.~K., Srivastava M.~K., Banerjee D.~P.~K., et al., 2021, ATel, 14882

Schmid, H.M. 1989, A\&A 211, L31

Shidatsu M., Negoro H., Mihara T., et al., 2021, ATel, 14846

Shore S.~N., Allen H., Bajer M., et al., 2021a, ATel, 14868

Shore S.~N., Teyssier F., Thizy O.,  2021b, ATel, 14881

Shore S.~N., Teyssier F., Guarro J., et al., 2021c, ATel, 14883

Sokolovsky K., Aydi E., Chomiuk L., et al., 2021, ATel, 14886

Taguchi K., Ueta, T., Isogai, K., 2021a, ATel, 14838

Taguchi K., Maheara H., Isogai K., et al., 2021b, ATel, 14858

Wagner, S.~J., HESS Collaboration, 2021a, ATel, 14844

Wagner, S.~J., HESS Collaboration, 2021b, ATel, 14857

Williams D., O'Brien T., Woudt P., et al., 2021, ATel, 14849

Woodward C.~E., Evans A., Banerjee D.~P.~K., et al., 2021a, ATel, 14866

Woodward C.~E., Banerjee D.~P.~K., Evans A., et al., 2021b, ATel, 14910

Zamanov R.~K., Stoyanov K.~A., Kostov A., et al., 2021a, ATel, 14974

Zamanov R.~K., Stoyanov K.~A., Nikolov Y.~M., et al., 2021b, arXiv:2109.11306

Zwitter T., Munari U., 2000, An introduction to analysis of spectra with IRAF, Univ. of Padova (2000iasd.book.....Z)

\end{document}